\documentstyle[twoside,fleqn,espcrc2,epsf]{article}
\begin{document}
\def\siml{{\ \lower-1.2pt\vbox{\hbox{\rlap{$<$}\lower6pt\vbox{\hbox{$\sim$}}}}\ }} 
\def\bfnabla{\mbox{\boldmath $\nabla$}}
\def\bfSigma{\mbox{\boldmath $\Sigma$}}
\def\bfsigma{\mbox{\boldmath $\sigma$}}
\def\als{\alpha_{\rm s}}
\def\al{\alpha}
\def\lQ{\Lambda_{\rm QCD}}
\def\vs{V^{(0)}_s}
\def\vo{V^{(0)}_o}
\newcommand{\ttbs}{\char'134}
\newcommand{\AmS}{{\protect\the\textfont2 A\kern-.1667em\lower.5ex\hbox{M}\kern-.125emS}}

\title{The heavy quark potential in pNRQCD}

\author{Antonio Vairo\address{Institut  f\"ur Hochenergiephysik, 
        \"Osterreichische Akademie der Wissenschaften\\ 
        Nikolsdorfergasse 18, A-1050 Vienna, Austria}
\thanks{FWF contract No. P12254; 'Acciones Integradas' 1999-2000, project No. 13/99.}}

\begin{abstract}
The heavy quarkonium static potential is discussed within the framework of potential NRQCD. 
Some quantitative statements are made in the kinematical situation $mv \gg \Lambda_{\rm QCD}$
at the level of accuracy of the next-to-leading order in the multipole expansion.
\end{abstract}

\maketitle

\section{INTRODUCTION}
In a recent series of papers \cite{BPSV1,BPSV2} a detailed study of 
a suitable Effective Field Theory for heavy quark bound states, called 
potential NRQCD (pNRQCD) \cite{pNRQCD}, has been started.  
Since several issues have been treated in that context, we address the reader 
to the quoted literature for a complete overview of the achieved results.  
Here we only mention the 1-loop matching \cite{pineda99} and the static 
energies of the hybrids \cite{brambilla99} presented in these proceedings. 
While in the following, due to its considerable importance,  
we recollect and summarize our understanding of the heavy quarkonium static potential.

Let us define, first, what we mean with heavy quarkonium potential. 
Being heavy quarkonium a non-relativistic bound system, it is characterized by at least 
three energy scales: the mass or hard scale, $m$, the momentum or soft scale, $m v$, 
corresponding to the inverse of the bound-state size 
and the energy or ultrasoft (US) scale, $m v^2$ ($v$ is the heavy quark velocity). 
As a consequence, when US degrees of freedom are neglected, 
heavy quarkonium can be described as a bound state $\phi$ 
governed by a non-relativistic Schr\"odinger equation of the type 
\begin{equation}
\left( {p^2\over m} + V({\bf r},{\bf p},{\bf S}_1,{\bf S}_2,m) \right)\phi =  E \phi.
\label{eqS}
\end{equation}
We will call $V$ the heavy quarkonium potential, 
which is in general a function of the quark distance ${\bf r}$, 
momentum ${\bf p}$, spin ${\bf S}_1$, ${\bf S}_2$, and mass $m$. 

Another relevant quantity for heavy quarkonium physics 
is expected to be the energy between static sources.  For heavy quarkonium 
in a singlet state this can be defined as
\begin{equation}
E_s(r) =  \displaystyle\lim_{T\to\infty} {i \over T} \ln  \langle W_\Box \rangle,
\label{eqW}
\end{equation}
where $W_\Box$ is the static Wilson loop of size ${\bf r} \times T$ and 
the symbol $\langle ~~ \rangle$ means the average over the gauge fields.
Often in the literature $E_s$ has been implicitly identified with the static limit 
of the Schr\"odinger potential, $\displaystyle\lim_{m\to\infty} V$. While it is reasonable to expect this 
identification to hold to some extent, there are no general grounds for it to be true in general. 
Indeed, already long time ago \cite{Appelquist} several doubts have been rised 
on the infrared consistency of that identification at least in perturbation theory.  
It is the goal of this contribution to make some quantitative statements  
on the difference $E_s - \displaystyle\lim_{m\to\infty} V$.  
This will be done in the last section. In the next section we perform, as an 
intermediate step, the matching in the singlet sector of the pNRQCD Lagrangian 
at the next-to-leading order in the multipole expansion.

\section{pNRQCD}
Another scale is relevant in QCD, the scale where nonperturbative 
effects start to become important. We will call this scale $\Lambda_{\rm QCD}$ 
and we will assume that $mv \gg \Lambda_{\rm QCD}$.
For sufficiently heavy quarkonium $v\ll 1$ and therefore 
the energy scales of the system are widely separated. This allows to systematically integrate out 
these scales by matching QCD with simpler but equivalent Effective Field Theories. 
The integration of the hard scale ($\sim m$)  gives rise to the effective theory known as 
non-relativistic QCD (NRQCD) \cite{NRQCD},  whereas the integration of the soft scale ($\sim mv$) 
gives rise to what we call potential NRQCD. 
Being $m$ and $mv$ well above $\Lambda_{\rm QCD}$ both matchings can be done perturbatively. 

By definition pNRQCD is the effective theory where only degrees of freedom 
below the soft scale remain dynamical. The surviving fields are quark-antiquark states  
(with US energy) and gluons with energy and momentum below $m v$.  
It is convenient to decompose the quark-antiquark states into singlets 
and octets under colour transformation. The relative coordinate ${\bf r}$, 
whose typical size is the inverse of the soft scale, is explicit and can be considered as small 
with respect to the remaining dynamical lengths in the system. Hence the gluon fields
can be systematically expanded in ${\bf r}$ (multipole expansion). Therefore the pNRQCD Lagrangian 
is constructed not only order by order in $1/m$, but also order by order in 
${\bf r}$. As a typical feature of an effective theory, all the non-analytic behaviour 
in ${\bf r}$ is encoded in the matching coefficients.

The most general pNRQCD Lagrangian density that can be constructed with these 
fields and that is compatible with the symmetries of NRQCD is given 
at order $1/m^0$ (but we write also explicitly the kinetic energy in the 
centre-of-mass frame) and at the leading order in the multipole expansion by:
\begin{eqnarray}
& &\hspace{-7mm}
{\cal L}_{\rm pNRQCD} =
{\rm Tr} \Biggl\{ {\rm S}^\dagger \left( i\partial_0 - {{\bf p}^2\over m} 
- V_s(r) + \dots  \right) {\rm S} 
\nonumber \\
& &+ {\rm O}^\dagger \left( iD_0 - {{\bf p}^2\over m} 
- V_o(r) + \dots  \right) {\rm O} \Biggr\}
\nonumber\\
& &  + g V_A ( r) {\rm Tr} \left\{  {\rm O}^\dagger {\bf r} \cdot {\bf E} \,{\rm S}
+ {\rm S}^\dagger {\bf r} \cdot {\bf E} \,{\rm O} \right\} 
\nonumber \\
& &   + g {V_B (r) \over 2} {\rm Tr} \left\{  {\rm O}^\dagger {\bf r} \cdot {\bf E} \, {\rm O} 
+ {\rm O}^\dagger {\rm O} {\bf r} \cdot {\bf E}  \right\},  
\label{pnrqcd0}
\end{eqnarray}
where ${\rm S} = {\rm S}({\bf r},{\bf R},t)$ and ${\rm O} = {\rm O}({\bf r},{\bf R},t)$ 
are the singlet and octet wave functions respectively and ${\bf R}$ 
is the centre-of-mass coordinate of the quark-antiquark system. 
All the gauge fields in Eq. (\ref{pnrqcd0}) are evaluated 
in ${\bf R}$ and $t$. In particular ${\bf E} \equiv {\bf E}({\bf R},t)$ and 
$iD_0 {\rm O} \equiv i \partial_0 {\rm O} - g [A_0({\bf R},t),{\rm O}]$. 

We call $V_s$ and $V_o$ the singlet and octet static matching potentials respectively.
By looking at the equations of motion of the Lagrangian (\ref{pnrqcd0}) it is clear 
that, as far as higher order terms in the multipole expansion 
(terms of order ${\bf r}$ or smaller in (\ref{pnrqcd0})) do not give 
potential-type contributions, $V_s$ and $V_o$ coincide with the static 
singlet and octet potential to be used in the heavy quarkonium Schr\"odinger equation. 
This happens  when the US scale $m v^2$ is the next relevant scale of the system 
(i.e. $\Lambda_{\rm QCD} \siml m v^2$). While, in the situation 
$m v \gg \Lambda_{\rm QCD} \gg mv^2$ one expects to have nonperturbative 
corrections to the static potential coming from higher order terms in the multipole 
expansion. Both situations will be discussed in the next section. 

Here we sketch the singlet matching at order $1/m^0$ and at the next-to-leading 
order in the multipole expansion. We refer the reader to \cite{BPSV1,BPSV2} for a 
complete and detailed discussion. 
The matching is in general done by comparing 2-fermion Green functions 
(plus external gluons at a scale below $m v$) in NRQCD and pNRQCD, order by 
order in $1/m$ and order by order in the multipole expansion.
In order to get the singlet potential, we choose the following Green function in NRQCD: 
\begin{equation}
I = \delta^3({\bf x}_1 - {\bf y}_1) \delta^3({\bf x}_2 - {\bf y}_2) \langle W_\Box \rangle , 
\label{vsnrqcd}
\end{equation}
where $W_\Box$ is the rectangular Wilson loop with edges $x_1 = (T/2,{\bf r}/2)$, $x_2 = (T/2,-{\bf r}/2)$, 
$y_1 = (-T/2,{\bf r}/2)$ and  $y_2 = (-T/2,-{\bf r}/2)$. 
In pNRQCD we obtain at the  next-to-leading order in the multipole expansion
\begin{eqnarray}
& &\hspace{-7mm}
I = Z_s(r) \delta^3({\bf x}_1 - {\bf y}_1) \delta^3({\bf x}_2 - {\bf y}_2) e^{-iTV_s(r)} 
\nonumber\\
& &\hspace{-7mm}
\times \! \Bigg(\! 1 -{ g^2 \over N_c} T_F V_A^2 (r)\int_{-T/2}^{T/2} \!\!\!\! dt 
\int_{-T/2}^{t} \!\!\!\! dt^\prime e^{-i(t-t^\prime)(V_o-V_s)} 
\nonumber\\
& & \hspace{4mm}
\times \langle {\bf r}\cdot {\bf E}^a(t) \phi(t,t^\prime)^{\rm adj}_{ab}
{\bf r}\cdot {\bf E}^b(t^\prime)\rangle \!\Bigg),
\label{vspnrqcdus}
\end{eqnarray}
where $\phi^{\rm adj}$ is a Schwinger (straight-line) string in the adjoint representation 
and fields with only temporal argument are evaluated in the centre-of-mass coordinate.
Comparing Eqs. (\ref{vsnrqcd}) and (\ref{vspnrqcdus}), one gets at the 
next-to-leading order in the multipole expansion the singlet wave-function normalization $Z_s$ 
and the singlet static potential $V_s$. $V_A$ and $V_o$ must have 
been previously obtained from the matching of suitable operators, but for the present purposes 
we only need the tree-level values: $V_A = 1$ and $V_o = (C_A/2-C_F)\alpha_{\rm s}/r$. 
Let us concentrate here on the matching potential $V_s$. 
By substituting the chromoelectric field correlator in Eq. (\ref{vspnrqcdus}) 
with its perturbative expression we obtain  at the next-to-leading order in the 
multipole expansion and at order $\alpha_{\rm s}^4 \ln \alpha_{\rm s}$
\begin{eqnarray}
& & V_s(r) = E_s(r)\big\vert_{\rm 2-loop+NNLL} 
\nonumber\\
& & \hspace{10mm} + C_F {\alpha_{\rm s}\over r} {\alpha^3_{\rm s}\over \pi}
{C_A^3\over 12} \ln {C_A \alpha_{\rm s} \over 2 r \mu},  
\label{vsu0}
\end{eqnarray}
where $E_s$ has been defined in Eq. (\ref{eqW}). We note that $V_s$ and $E_s$ would coincide 
in QED and that therefore the effect we are studying here is a genuine QCD feature. 
The 2-loop contribution to $E_s$ has been calculated in \cite{twoloop}. The NNLL contributions 
arise from the diagrams studied first in \cite{Appelquist} and shown below. An explicit calculations gives 
\begin{figure}[h]
\makebox[0cm]{\phantom b}
\put(0,0){\epsfxsize=4truecm \epsfbox{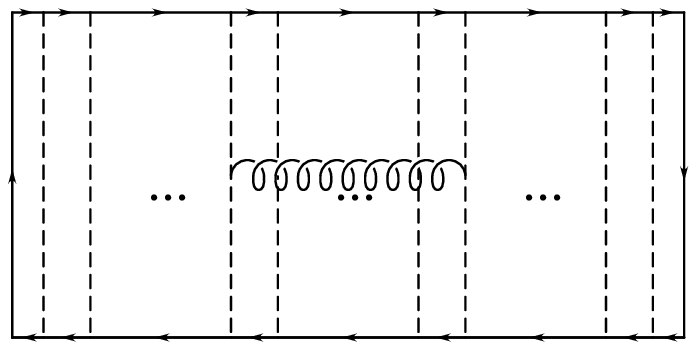}}
\put(115,25){$= -\displaystyle{C_FC_A^3 \alpha_{\rm s}^4 \over 12 \pi r} 
\ln {C_A \alpha_{\rm s}\over 2}$}
\put(130,5){$+ \displaystyle O\left( 1/T \right)$}
\put(0,-40){Inserting this and the 2-loop contribution in Eq.}
\put(0,-53){(\ref{vsu0}) we get}
\vspace{-8mm}
\end{figure}
\begin{eqnarray}
& & \hspace{-7mm} 
V_s(r) \equiv  - C_F {\alpha_{V}(r,\mu) \over r}, \nonumber\\
& & \hspace{-7mm} 
{\alpha}_{V}(r, \mu)=\alpha_{\rm s}(r)
\left\{1+\left(a_1+ 2 {\gamma_E \beta_0}\right) {\alpha_{\rm s}(r) \over 4\pi}\right. \nonumber\\
& &\hspace{-5mm} 
+{\alpha_{\rm s}^2(r) \over 16\,\pi^2}
\bigg[\gamma_E\left(4 a_1\beta_0+ 2{\beta_1}\right)+\left( {\pi^2 \over 3}+4 \gamma_E^2\right) 
{\beta_0^2}
\nonumber\\
& &\hspace{12mm} 
+ a_2\bigg] \left. + {C_A^3 \over 12}{\alpha_{\rm s}^3(r) \over \pi} \ln{ r \mu}\right\},
\label{newpot}
\end{eqnarray}
where $\beta_n$ are the coefficients of the beta function ($\alpha_{\rm s}$ is in the $\overline{\rm MS}$ scheme), 
and $a_1$ and $a_2$ are given in \cite{twoloop}. 

We conclude this section noticing that the octet matching potential $V_o$ can be calculated 
in the same way as done for the singlet. In this case the relevant NRQCD Green function 
could be chosen to be $\delta^3({\bf x}_1 - {\bf y}_1) \delta^3({\bf x}_2 - {\bf y}_2) \langle T^a W_\Box T^a\rangle$, 
where the colour matrices are inserted in the endpoint Schwinger strings.  
Even if this Green function is gauge dependent, as discussed in \cite{BPSV2}, the matching 
should guarantee a gauge invariant definition of $V_o$. A 2-loop calculation is still not available, 
but work is in progress \cite{schroeder}.

\section{THE STATIC SINGLET POTENTIAL}
In the previous section we have established the connection between $E_s$, the energy of static 
sources and $V_s$ the  singlet static matching potential of pNRQCD. Here we discuss in two different 
kinematical situations the connection of $V_s$ with the static limit of the heavy quarkonium potential 
defined through the Schr\"odinger equation. 

A) $\Lambda_{\rm QCD}\siml mv^2$. This situation is expected to hold for toponium and for 
the bottomonium (charmonium?) ground state. As already mentioned, in this situation $V_s$, as given by 
Eq. (\ref{newpot}), {\it is} the heavy quarkonium static potential in the sense given 
in the introduction. The explicit $\mu$ dependence of it originates from the fact that the US 
degrees of freedom (which have the same scale of the kinetic energy and therefore do not belong to 
the potential) have been explicitly subtracted out form the static Wilson loop. 
This fact is not surprising if we understand the heavy quarkonium potential 
as a matching coefficient of pNRQCD. As a consequence even in a purely perturbative 
regime the static heavy quarkonium potential (as well as $\alpha_V$) 
turns out to be an infrared sensitive quantity. 
In this situation nonperturbative effects are only of non-potential nature
(see for instance the Leutwyler--Voloshin type corrections in the situation 
$\Lambda_{\rm QCD}\ll mv^2$ \cite{Voloshin}). 
Finally we mention the quite obvious fact that, when calculating any physical observable, the $\mu$ 
dependence in (\ref{newpot}) must cancel against $\mu$-dependent contributions 
coming from the US gluons (see for instance \cite{KP}). 

B) $m v \gg \Lambda_{\rm QCD} \gg mv^2$.  Since in this situation there is a physical scale ($\Lambda_{\rm QCD}$)
above the US scale, a potential can be properly defined only once this scale has been integrated out.
At the next-to-leading order in the multipole expansion we get 
\begin{eqnarray}
& & \hspace{-7mm} 
V(r) = -C_F {\alpha_{V}(r,\mu)\over  r }
\nonumber\\
& &  \hspace{1mm} 
 -i{g^2 \over N_c}T_F V_A^2(r){r^2\over 3} \int_0^\infty \!\! dt \, 
e^{-it(V_o-V_s)} 
\nonumber\\
& &  \hspace{18mm} 
\times \langle {{\bf E}^a}(t)\phi(t,0)^{\rm adj}_{ab}{\bf E}^{b}(0) \rangle(\mu),
\label{vsnp}\\
& &\nonumber \\
& & \hspace{-7mm} 
e^{-it(V_o-V_s)} \! = \! 1 -it(V_o-V_s) - {t^2\over 2}(V_o-V_s)^2 + \dots 
\nonumber
\end{eqnarray}
Therefore, the heavy quarkonium static potential $V$ is given in this situation by the sum of the purely 
perturbative piece calculated in Eq. (\ref{newpot}) and a new term carrying   
also nonperturbative contributions (contained into non-local gluon field correlators). 
This last one can be organized as a series of power of $r^n$. 
We stress that due to the condition $m v \gg \Lambda_{\rm QCD}$
this expansion makes sense only in the short-range. Typically the nonperturbative piece 
of Eq. (\ref{vsnp}) absorbs the $\mu$ dependence of $\alpha_V$ (see \cite{BPSV2} for an example) 
so that the resulting potential $V$ is now scale independent.\vspace{4mm} 

The infrared sensitivity of the static potential can also be expressed in terms  
of renormalons (see for instance \cite{renormalon}). Rephrasing them in the 
Effective Field Theory language of pNRQCD we can say that the singlet matching potential 
$V_s$, as defined in Eq. (\ref{newpot}), suffers from IR 
renormalons ambiguities with the following structure
\begin{equation}
V_s(r) \vert_{\rm IR\, ren} = C_0 + C_2 r^2 + \dots
\label{ren1}
\end{equation}
The constant $C_0 \sim \Lambda_{\rm QCD}$ is known to be cancelled by the IR pole 
mass renormalon ($2 m_{\rm pole}\vert_{\rm IR\, ren} = - C_0$, \cite{thesis}). 
While Eq. (\ref{vsnp}) provides us with the explicit expression for the 
operator which absorbs the $C_2 \sim \Lambda_{\rm QCD}^3$ ambiguity \cite{BPSV2}. 
More precisely the order $r^2$ term on the right-hand side of Eq. (\ref{vsnp}) suffers 
from UV and IR renormalons. The UV renormalon ambiguity of it (which can be calculated 
simply by substituting the chromoelectric field correlator with its perturbative expression 
and summing up the leading log of all the bubble diagrams) cancels exactly the second term 
in the expansion (\ref{ren1}):  
\begin{eqnarray}
& & \hspace{-7mm} 
-i{g^2 \over N_c}T_F V_A^2(r){r^2\over 3} 
\label{ren2}\\
& & \hspace{-7mm}
\times \!\! \int_0^\infty \!\!\! dt \langle {{\bf E}^a}(t)\phi(t,0)^{\rm adj}_{ab}{\bf E}^{b}(0) \rangle(\mu)
\bigg\vert_{\rm UV\, ren} = - C_2 r^2.
\nonumber\\
\nonumber
\end{eqnarray}

{\bf Acknowledgements}

I thank Nora Brambilla, Antonio Pineda and Joan Soto for collaboration on the work presented here.

\end{document}